MPI-PhT/95-40
DESY 95-088

hep-lat/9505011 12 May 1995

# Computation of the relation between the bare lattice coupling and the $\overline{\mathrm{MS}}$ coupling in SU($N$) gauge theories to two loops

Martin Lüscher

Deutsches Elektronen-Synchrotron DESY
Notkestrasse 85, D-22603 Hamburg, Germany

Peter Weisz

Max-Planck-Institut für Physik
Föhringer Ring 6, D-80805 München, Germany

**Abstract**

The perturbation expansion of the $\overline{\mathrm{MS}}$ coupling in the SU($N$) gauge theory in powers of the bare lattice coupling is extended to two-loop order. A discussion of our results has already been published elsewhere. We here describe the calculation in some detail. It relies on the background field technique and a recently introduced coordinate space method to evaluate lattice Feynman integrals.

May 1995

## 1. Introduction

Our aim in this paper is to compute the coefficient $d_2(s)$ in the expansion

$$\alpha_{\overline{\rm MS}}(s/a) = \alpha_0 + d_1(s)\,\alpha_0^2 + d_2(s)\,\alpha_0^3 + \ldots, \tag{1.1}$$

which relates the running coupling in the $\overline{\rm MS}$ scheme of dimensional regularization to the bare lattice coupling $\alpha_0$ in the pure SU($N$) gauge theory. The motivation for this calculation and references to earlier work were given in a previous publication [1]. We assume the reader is familiar with this paper and also refs.[2,3], where we discuss the background field formalism in lattice gauge theory and describe some coordinate space techniques to evaluate lattice Feynman diagrams.

In sect. 2 the strategy of the computation is outlined. After summarizing some results for the dimensionally regularized theory in sect. 3, we write down the Feynman rules for the lattice theory in the presence of a background field (sect. 4). We then first calculate the one-loop diagrams contributing to the background field 2-point function (sect. 5). The computation of the two-loop diagrams is discussed in some detail in sect. 6. We finally collect our results and obtain $d_2(s)$ (sect. 7).

## 2. Background field technique and renormalization constants

The probably most effective way to compute the renormalization constants required to match different regularization schemes is based on the background field technique. The method has previously been employed to calculate the one-loop coefficient $d_1(s)$ [4–6] and its applicability to lattice gauge theory at all orders of perturbation theory has been established in ref.[2]. We have thus decided to make use of this technique to compute $d_2(s)$. In the following the notation is as in ref.[2].

Let $\Gamma^{(j,k,l)}$ be the bare vertex functions in the lattice theory and $\Gamma_{\rm R}^{(j,k,l)}$ the renormalized vertex functions in the $\overline{\rm MS}$ scheme of dimensional regularization at $\epsilon = 0$. The lattice vertex functions depend on the bare coupling $g_0$, the bare gauge parameter $\lambda_0$ and the lattice spacing $a$. In the case of the $\overline{\rm MS}$ vertex functions the parameters are the renormalized coupling $g$, the renormalized gauge parameter $\lambda$ and the normalization mass $\mu$.

As discussed in ref.[2] the vertex functions $\Gamma^{(j,k,l)}$ and $\Gamma_{\rm R}^{(j,k,l)}$ are related by

$$\Gamma_{\rm R}^{(j,k,l)} = \mathcal{Z}_3^{k/2} \tilde{\mathcal{Z}}_3^{l} \, \Gamma^{(j,k,l)} + {\rm O}(a), \tag{2.1}$$

where, for the parameters $g$ and $\lambda$ on the left hand side, one should substitute

$$g = \mathcal{Z}_1^{-1} \mathcal{Z}_3^{3/2} g_0, \tag{2.2}$$

$$\lambda = \mathcal{Z}_3 \lambda_0. \tag{2.3}$$

The renormalization constants $\mathcal{Z}_1$, $\mathcal{Z}_3$ and $\tilde{\mathcal{Z}}_3$ are formal power series in $g_0$ with coefficients depending on $\lambda_0$ and $a\mu$.

The couplings appearing in eq.(1.1) are $\alpha_{\overline{\rm MS}}(\mu) = g^2/4\pi$ and $\alpha_0 = g_0^2/4\pi$. So if we define the coefficients $\mathcal{Z}^{(l)}$ through

$$\mathcal{Z}_1^2 \mathcal{Z}_3^{-3} = 1 + \sum_{l=1}^{\infty} \mathcal{Z}^{(l)} g_0^{2l}, \tag{2.4}$$

it follows that

$$d_1(s) = -4\pi \; \mathcal{Z}^{(1)} \Big|_{\mu = s/a}, \tag{2.5}$$

$$d_2(s) = d_1(s)^2 - (4\pi)^2 \; \mathcal{Z}^{(2)} \Big|_{\mu = s/a}. \tag{2.6}$$

An important point to note here is that the coefficients $\mathcal{Z}^{(l)}$ (and hence the $d_l(s)$) are independent of the gauge parameter.

To calculate $\mathcal{Z}^{(1)}$ and $\mathcal{Z}^{(2)}$ we consider the background field 2-point function

$$\Gamma_{\rm R}^{(2,0,0)}(p,-p)_{\mu\nu}^{ab} = -\delta^{ab}(\delta_{\mu\nu}p^2 - p_\mu p_\nu)[1 - \nu_{\rm R}(p)]/g^2 \tag{2.7}$$

and the inverse gluon propagator

$$\Gamma_{\rm R}^{(0,2,0)}(p,-p)_{\mu\nu}^{ab} = -\delta^{ab}\{(\delta_{\mu\nu}p^2 - p_\mu p_\nu)[1 - \omega_{\rm R}(p)] + \lambda p_\mu p_\nu\}. \tag{2.8}$$

The tensor structure of these functions, as given above, is implied by the symmetries of the theory. The Lorentz invariant amplitudes $\nu_{\rm R}(p)$ and $\omega_{\rm R}(p)$ are of order $g^2$. On the lattice we define the analogous amplitudes $\nu(p)$ and

$\omega(p)$ through

$$\sum_{\mu} \Gamma^{(2,0,0)}(p,-p)^{ab}_{\mu\mu} = -\delta^{ab} 3\hat{p}^2 [1-\nu(p)]/g_0^2, \tag{2.9}$$

$$\sum_{\mu} \Gamma^{(0,2,0)}(p,-p)^{ab}_{\mu\mu} = -\delta^{ab} \hat{p}^2 \{3[1-\omega(p)]+\lambda_0\}, \tag{2.10}$$

where $\hat{p}_\mu = (2/a)\sin(ap_\mu/2)$. They are invariant under lattice rotations and reflections, but not under continuous Lorentz transformations.

In the rest of this section all contributions to $\nu(p)$ and $\omega(p)$ which vanish in the continuum limit $a \to 0$ are neglected. From eqs.(2.1)–(2.3) we then infer that

$$\mathcal{Z}_1^2 \mathcal{Z}_3^{-3} = [1-\nu(p)] \,/\, [1-\nu_{\rm R}(p)], \tag{2.11}$$

$$\mathcal{Z}_3 = [1-\omega_{\rm R}(p)] \,/\, [1-\omega(p)]. \tag{2.12}$$

Note that $\mathcal{Z}_1$ and $\mathcal{Z}_3$ are only implicitly determined by these equations, since $\nu_{\rm R}(p)$ and $\omega_{\rm R}(p)$ are functions of $g$ and $\lambda$, which in turn are given by eqs.(2.2) and (2.3). Introducing the loop expansions

$$\nu_{\rm R}(p) = \sum_{l=1}^{\infty} g^{2l} \nu_{\rm R}^{(l)}(p), \tag{2.13}$$

$$\omega_{\rm R}(p) = \sum_{l=1}^{\infty} g^{2l} \omega_{\rm R}^{(l)}(p), \tag{2.14}$$

and similarly for the lattice amplitudes,

$$\nu(p) = \sum_{l=1}^{\infty} g_0^{2l} \nu^{(l)}(p), \tag{2.15}$$

$$\omega(p) = \sum_{l=1}^{\infty} g_0^{2l} \omega^{(l)}(p), \tag{2.16}$$

it follows that

$$\mathcal{Z}^{(1)} = \left\{ \nu_{\rm R}^{(1)}(p) - \nu^{(1)}(p) \right\}_{\lambda=\lambda_0}, \tag{2.17}$$

$$\mathcal{Z}_3^{(1)} = \left\{\omega^{(1)}(p) - \omega_{\rm R}^{(1)}(p)\right\}_{\lambda=\lambda_0}. \tag{2.18}$$

At two-loop order the renormalization of the gauge parameter must be taken into account and one obtains

$$\mathcal{Z}^{(2)} = \left\{\nu_{\rm R}^{(2)}(p) - \nu^{(2)}(p) + \lambda\frac{\partial\nu_{\rm R}^{(1)}(p)}{\partial\lambda}\,\mathcal{Z}_3^{(1)}\right\}_{\lambda=\lambda_0}. \tag{2.19}$$

It is now clear that to compute $d_2(s)$ it suffices to work out $\omega_{\rm R}(p)$ and $\omega(p)$ to one-loop order and $\nu_{\rm R}(p)$ and $\nu(p)$ to two-loop order. While $\nu_{\rm R}^{(1)}(p)$ is required for general $\lambda$, we may choose the Feynman gauge $\lambda = \lambda_0 = 1$ in all other cases. The computations are significantly simplified by this choice, especially in the lattice theory.

## 3. Computation of $\omega_R$ and $\nu_R$

Starting from the total action, eq.(2.6) of ref.[2], it is straightforward to deduce the Feynman rules for the dimensionally regularized theory with background field. The relevant diagrams have been listed in refs.[7,8], for example. The computation of these diagrams is standard so that here we only quote the final results in the $\overline{\rm MS}$ scheme [9,10].

At one-loop order the background field propagator is given by

$$\nu_{\rm R}^{(1)}(p) = \frac{N}{16\pi^2}\left\{\frac{11}{3}\ln(\mu^2/p^2) + \frac{205}{36} + \frac{3}{2\lambda} + \frac{1}{4\lambda^2}\right\}, \tag{3.1}$$

while for the gluon self-energy one obtains

$$\omega_{\rm R}^{(1)}(p) = \frac{N}{16\pi^2}\left\{\left(\frac{13}{6} - \frac{1}{2\lambda}\right)\ln(\mu^2/p^2) + \frac{97}{36} + \frac{1}{2\lambda} + \frac{1}{4\lambda^2}\right\}. \tag{3.2}$$

$\nu_R(p)$ has previously been computed to two loops by Ellis [11]. His result was checked independently by van de Ven [12] and by one of the present authors. Prior to the calculation of Ellis, the divergent parts of the diagrams (from which the two-loop coefficient of the Callan-Symanzik $\beta$–function can be extracted) had been computed in refs.[7,8]. In the Feynman gauge the result is

$$\nu_{\rm R}^{(2)}(p)\Big|_{\lambda=1} = \left(\frac{N}{16\pi^2}\right)^2 \left\{8\ln(\mu^2/p^2) - 6\zeta(3) + \frac{577}{18}\right\}, \qquad (3.3)$$

where $\zeta(z)$ denotes Riemann's zeta function (ref.[13], §9.522).

## 4. Lattice perturbation theory

In the remainder of this paper we discuss the computation of the lattice amplitudes $\nu(p)$ and $\omega(p)$. We begin by describing the Feynman rules for the lattice theory, starting from the total action given in subsect. 5.1 of ref.[2]. For notational convenience we now set $a = 1$.

### *4.1 Vertices*

The total action $S_{\rm tot}$ of the lattice theory can be expanded in powers of its arguments according to

$$\begin{aligned} S_{\rm tot}[B,q,\bar{c},c] = \sum_{j,k=0}^{\infty}\sum_{l=0,1}\frac{1}{j!\,k!}\int_{-\pi}^{\pi}\frac{{\rm d}^4p_1}{(2\pi)^4}\ldots\frac{{\rm d}^4s_l}{(2\pi)^4} \\ \times (2\pi)^4\delta_P(p_1+\ldots+s_l)\,V^{(j,k,l)}(p_1,\ldots,s_l)^{a_1\ldots d_l}_{\mu_1\ldots\nu_k} \\ \times \tilde{B}^{a_1}_{\mu_1}(-p_1)\ldots\tilde{B}^{a_j}_{\mu_j}(-p_j)\,\tilde{q}^{b_1}_{\nu_1}(-q_1)\ldots\tilde{q}^{b_k}_{\nu_k}(-q_k) \\ \times \tilde{\bar{c}}^{c_1}(-r_1)\ldots\tilde{\bar{c}}^{c_l}(-r_l)\,\tilde{c}^{d_1}(-s_1)\ldots\tilde{c}^{d_l}(-s_l) \end{aligned} \qquad (4.1)$$

(for unexplained notations see ref.[2]). The vertices $V^{(j,k,l)}$ are defined through this expansion and the requirement that they should be invariant under permutations of the momenta and indices associated with the same type of fields.

A further set of vertices arises from the a priori measure in the functional integral. The jacobian appearing in eq.(5.10) of ref.[2] is equivalent to the "measure" action

$$S_{\rm m}[q] = -\sum_{x\in\Lambda}\sum_{\mu=0}^{3}{\rm tr}\,\left\{\ln\left[J\left(g_0q_\mu(x)\right)\right]\right\}. \qquad (4.2)$$

The associated vertices, $V_{\mathrm{m}}^{(k)}$, are obtained by expanding

$$S_{\mathrm{m}}[q] = \sum_{k=2}^{\infty} \frac{1}{k!} \int_{-\pi}^{\pi} \frac{\mathrm{d}^4 p_1}{(2\pi)^4} \dots \frac{\mathrm{d}^4 p_k}{(2\pi)^4}$$

$$\times (2\pi)^4 \delta_P(p_1 + \dots + p_k) \, V_{\mathrm{m}}^{(k)}(p_1, \dots, p_k)_{\mu_1 \dots \mu_k}^{a_1 \dots a_k}$$

$$\times \, \tilde{q}_{\mu_1}^{a_1}(-p_1) \dots \tilde{q}_{\mu_k}^{a_k}(-p_k). \tag{4.3}$$

Note that $V_{\mathrm{m}}^{(k)}$ is proportional to $g_0^k$ and so is of higher order than the vertex $V^{(0,k,0)}$. In particular, the measure vertices do not contribute at tree-level of perturbation theory.

For small values of $j+k+2l$ it is not difficult to obtain explicit expressions for the vertices $V^{(j,k,l)}$ by hand. For illustration the complete set of 2- and 3-point vertices is listed in appendix A. In our two-loop calculations vertices with up to 6 legs occur. The formulae for these are rather lengthy and it is advisable to generate them using an algebraic manipulation program. A possible way to proceed has been described some time ago in ref.[14]. Following these lines we have written MAPLE programs for all required vertices. The programs deliver an abstract algebraic expression in the colour tensors tr $\{T^{a_1} \dots T^{a_n}\}$, the generalized Kronecker delta symbols

$$\delta_{\mu_1 \dots \mu_n} = \begin{cases} 1 & \text{if } \mu_1 = \mu_2 = \dots = \mu_n, \\ 0 & \text{otherwise,} \end{cases} \tag{4.4}$$

and the sines and cosines of the momenta.

*4.2 Graphical representation*

The Feynman rules for the vertex functions $\Gamma^{(j,k,l)}$ can now be set up in the usual way. It may be helpful at this point to consider the diagrams contributing to $\Gamma^{(2,0,0)}$ at one- and two-loop order shown in fig. 1 in sect. 5 and fig. 2 in sect. 6. There are two types of internal lines, representing the propagation of the gluon field $q_\mu$ (wiggly lines) and the ghost fields $c$ and $\bar{c}$ (directed dashed lines). The propagators associated with these lines are given by

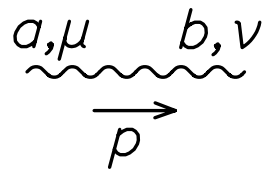

$$= \delta^{ab}\{\delta_{\mu\nu} + (\lambda_0^{-1} - 1)\hat{p}_\mu \hat{p}_\nu / \hat{p}^2\} / \hat{p}^2, \tag{4.5}$$

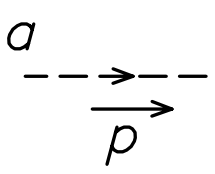

$$= \delta^{ab}/\hat{p}^2, \tag{4.6}$$

where the notation $\hat{p}_\mu = 2\sin(\frac{1}{2}p_\mu)$ has been used (cf. appendix A). There are no factors associated with external gluon and ghost lines for which the obvious graphical notation is employed.

Background field external lines are denoted by a wiggly line with a terminating cross. In particular, at tree-level the background field 2-point function is given by the diagram

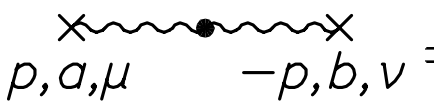

$$= -V^{(2,0,0)}(p,-p)^{ab}_{\mu\nu}. \tag{4.7}$$

The vertices $V^{(j,k,l)}$ are represented by filled circles from which $j+k+2l$ lines emerge. The $j$ background field lines are external and thus terminate with a cross. The notation is best illustrated by considering the 3-point vertices

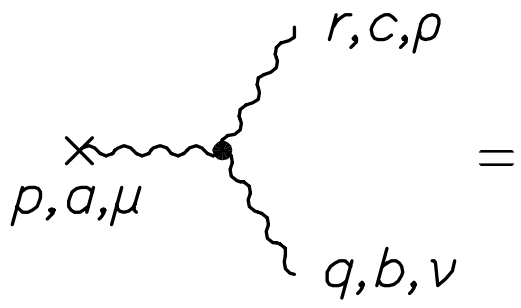

$$= -V^{(1,2,0)}(p,q,r)^{abc}_{\mu\nu\rho}, \tag{4.8}$$

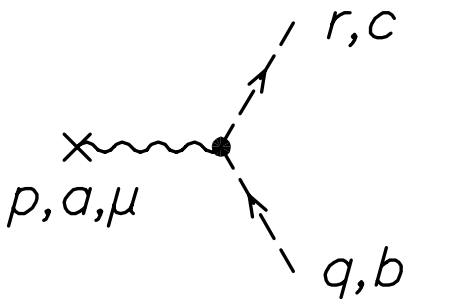

$$= -V^{(1,0,1)}(p,q,r)^{abc}_{\mu}. \tag{4.9}$$

The graphical symbol for the measure vertices $V^{(k)}_{\mathrm{m}}$ is an open square as in

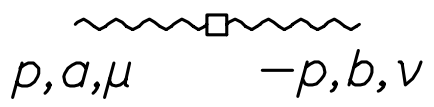

$$= -V^{(2)}_{\mathrm{m}}(p,-p)^{ab}_{\mu\nu}. \tag{4.10}$$

In all these figures the momenta associated with the legs of the vertices are in-going momenta. Note that it is always minus the vertex $V^{(j,k,l)}$ or $V^{(k)}_{\mathrm{m}}$ which is associated with the corresponding graphical element.

Without loss we may assign external and loop momenta in such a way that momentum is exactly conserved at each vertex (not just modulo $2\pi$). The loop momenta $k_1,\ldots,k_n$ must be integrated over the Brillouin zone with measure $\mathrm{d}^4k_1\ldots\mathrm{d}^4k_n/(2\pi)^{4n}$. Otherwise the rules are as in the continuum theory.

*4.3 Computation of Feynman integrands*

At two-loop order the integrands associated with the diagrams contributing to the background field 2-point function may easily involve several hundred terms. To compute the integrands and to be able to manipulate them safely a fully automated approach is thus suggested. The algebraic calculation of Feynman integrands on a computer is standard in continuum theories. Similar programs may be used for lattice diagrams to work out the required contractions of colour and Lorentz indices. In particular, the contraction of the colour indices may be carried out by repeated application of the identities

$$\mathrm{tr}\,\{T^a X T^a Y\} = \frac{1}{2N}\,\mathrm{tr}\,\{XY\} - \frac{1}{2}\,\mathrm{tr}\,\{X\}\;\mathrm{tr}\,\{Y\}, \tag{4.11}$$

$$\mathrm{tr}\,\{T^a X\}\;\mathrm{tr}\,\{T^a Y\} = \frac{1}{2N}\,\mathrm{tr}\,\{X\}\;\mathrm{tr}\,\{Y\} - \frac{1}{2}\,\mathrm{tr}\,\{XY\}, \tag{4.12}$$

which hold for arbitrary complex $N \times N$ matrices $X$ and $Y$.

Concerning the contraction of Lorentz indices we note that the lattice symmetries are not nearly as restrictive as the continuous Lorentz invariance. This is reflected by the presence of the generalized Kronecker delta symbols (4.4) in the algebraic expressions for the vertices. In general the result of the computation is hence not simply composed of scalar products of Lorentz vectors. One rather ends up with products of sines and cosines of various linear combinations of the momenta, with indices that are summed over or are equal to one of the external indices. To achieve a uniform representation of such terms it is sometimes useful to expand the products in sums of sines and cosines. The latter may then again be factorized into sums of products of

$$\hat{k}_\mu^2 = 2 - 2\cos(k_\mu) \quad \text{and} \quad \mathring{k}_\mu = \sin(k_\mu), \tag{4.13}$$

where $k$ is any of the basic momenta. Since

$$\mathring{k}_\mu^2 = \hat{k}_\mu^2 - \tfrac{1}{4}\hat{k}_\mu^4, \tag{4.14}$$

we may insist that there is at most one factor $\mathring{k}_\mu$ in each product, for any given momentum $k$ and index $\mu$.

It should be quite clear from these remarks that the available programs to compute Feynman integrands in continuum theories are not directly applicable to the lattice theory. We have thus developed our own libraries of MAPLE programs, which enable us to obtain the integrand associated with any given lattice Feynman diagram without manual calculations. One only needs to supply the diagram data in some appropriate form and to indicate which indices are to be contracted. Similar sets of programs have recently been described in refs.[15,16].

## 5. Computation of $\nu^{(1)}(p)$ and $\omega^{(1)}(p)$

At one-loop order there are 4 diagrams contributing to $\Gamma^{(2,0,0)}(p,-p)^{ab}_{\mu\nu}$ (see fig. 1). Accordingly we have

$$\nu^{(1)}(p)\Big|_{\lambda_0=1} = \sum_{i=1}^{4} \nu_i(p), \tag{5.1}$$

where the label $i$ counts the diagrams. In the case of the "tadpole" diagrams, $i=3,4$, the corresponding Feynman integrals are of the form

$$\hat{p}^2\nu_i(p) = \int_{-\pi}^{\pi} \frac{\mathrm{d}^4k}{(2\pi)^4}\,\frac{F_i(p,k)}{\hat{k}^2}, \tag{5.2}$$

while for $i=1,2$ the loop momentum may be chosen such that

$$\hat{p}^2\nu_i(p) = \int_{-\pi}^{\pi} \frac{\mathrm{d}^4k}{(2\pi)^4}\,\frac{F_i(p,k)}{\hat{k}_+^2\hat{k}_-^2}, \qquad k_\pm = k \pm \tfrac{1}{2}p. \tag{5.3}$$

The functions $F_i(p,k)$ are some complicated polynomials in the sines and cosines of the momenta. They are invariant under the lattice symmetries and we may assume that $F_i(p,k)=F_i(p,-k)$ since the other factors in the integrals are invariant under reflections of $k$.

We are interested in evaluating the diagrams in the continuum limit. Since the lattice spacing $a$ has been set to 1, the limit amounts to taking $p\to 0$. Our task then is to work out the asymptotic expansion of the integrals up to order $p^2$. In the conventional language the terms of order $p^0$ represent a quadratically divergent contribution. These terms cancel in the sum of all diagrams. Terms of order $p^1$ (which would represent a linearly divergent contribution) are excluded by symmetry. In the continuum limit we are then left with the terms of order $p^2$.

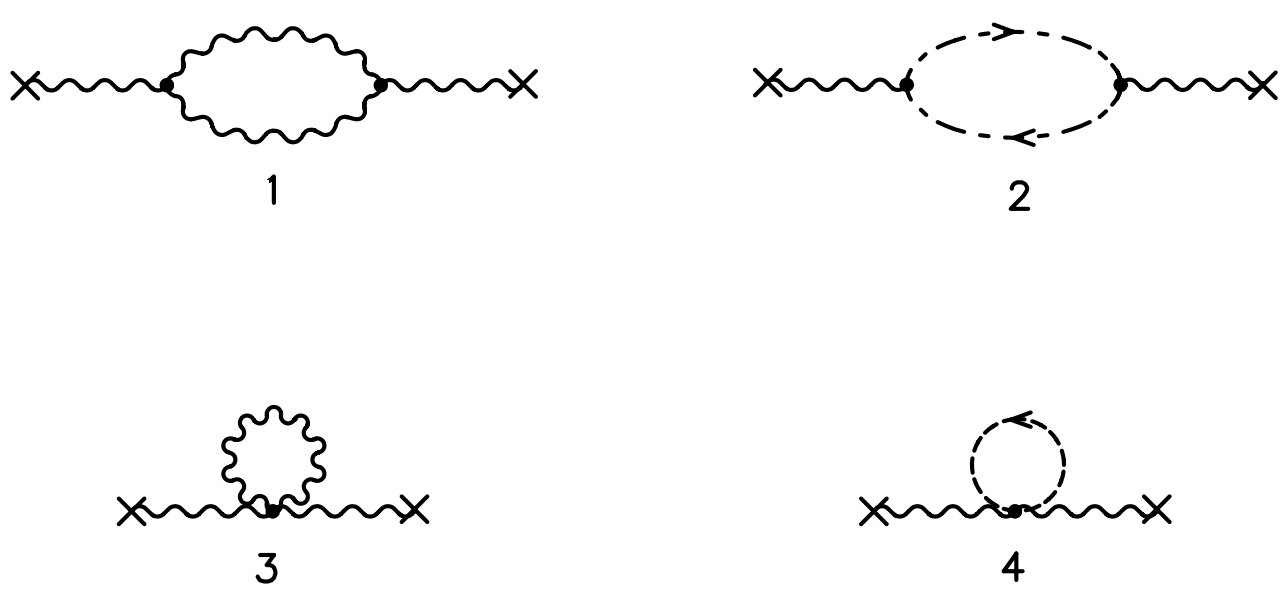

Fig. 1. One-loop diagrams contributing to the background field 2-point function

To compute the asymptotic form of the integrals we first expand the numerators $F_i(p,k)$ of the integrands in powers of $p$. It suffices to keep all terms up to second order, since the remainder makes a contribution of order $p^3$ (or higher) to $\hat{p}^2\nu_i(p)$ and so is negligible in the continuum limit. The numerators of the integrands then are of the form

$$f(k) + \sum_{\mu,\nu=0}^{3} p_\mu p_\nu f_{\mu\nu}(k) \tag{5.4}$$

with coefficients $f(k)$ and $f_{\mu\nu}(k)$ that transform like a scalar and a symmetric tensor under the lattice symmetries.

In the case of the tadpole diagrams, $i = 3,4$, the symmetries allow us to substitute

$$\sum_{\mu,\nu=0}^{3} p_\mu p_\nu f_{\mu\nu}(k) \rightarrow \tfrac{1}{4}p^2 \sum_{\mu=0}^{3} f_{\mu\mu}(k). \tag{5.5}$$

Inspection then shows that the integrals reduce to a linear combination of 1 and the integral (B.1) introduced in appendix B.

To evaluate the other diagrams, $i = 1,2$, we note that the integrals with integrands proportional to $f(k)$ reduce to linear combinations of the integrals (B.10),(B.13) and (B.14). In the integrals with integrands proportional to $f_{\mu\nu}(k)$ we substitute

$$f_{\mu\nu}(k) = c\,\delta_{\mu\nu} + f'_{\mu\nu}(k), \tag{5.6}$$

where $c$ is a constant and $f'_{\mu\nu}(k)$ of order $k^2$ for $k \rightarrow 0$. The terms proportional to $c$ reduce to the integral $A(p)$ [eq.(B.8)]. The remaining terms are sufficiently regular that we can take the continuum limit of the integrals simply by setting $p = 0$ in the denominator of the integrand. Taking the lattice symmetries into account these integrals reduce to a linear combination of 1 and the integrals (B.1) and (B.5).

The steps outlined above are sufficiently systematic to be programmable. We have in fact written a set of programs which perform the necessary expansions and separation of terms. The integration, which amounts to matching the terms with the integrals listed in appendix B, has also been automated. The programs thus perform the complete computation. At no point does one need to copy and manipulate the terms by hand.

Our final result then is

$$\nu^{(1)}(p)\Big|_{\lambda_0=1} = \frac{N}{16\pi^2}\left\{-\frac{11}{3}\ln(p^2) + \frac{64}{9}\right\} - \frac{1}{8N} + N\left\{\frac{5}{36}P_1 + \frac{11}{3}P_2 + \frac{1}{16}\right\} + \mathrm{O}(p^2), \tag{5.7}$$

where the constants $P_1$ and $P_2$ are defined in appendix B.

The gluon self-energy $\omega^{(1)}(p)$ has previously been calculated in ref.[17]. Apart from the diagram (4.10) which involves the measure vertex $V_{\mathrm{m}}^{(2)}$, all diagrams contributing to the self-energy are obtained from those in fig. 1 by replacing the background field external lines by gluon external lines. The computation of the corresponding Feynman integrals proceeds as in the case of the background field 2-point function. We only need the result in the Feynman gauge and there it is given by

$$\omega^{(1)}(p)\Big|_{\lambda_0=1} = \frac{N}{16\pi^2}\left\{-\frac{5}{3}\ln(p^2) + \frac{28}{9}\right\} - \frac{1}{8N} + N\left\{\frac{7}{72}P_1 + \frac{5}{3}P_2 + \frac{1}{16}\right\} + \mathrm{O}(p^2). \tag{5.8}$$

As an additional check of our programs we did in fact compute the full amplitudes $\Gamma^{(2,0,0)}(p,-p)^{ab}_{\mu\nu}$ and $\Gamma^{(0,2,0)}(p,-p)^{ab}_{\mu\nu}$ in an arbitrary gauge and verified that the calculated functions satisfied the Ward identities.

## 6. Computation of $\nu^{(2)}(p)$

We now describe the computation of the background field 2-point function at two-loop order in some detail. This is the most difficult part of our calculations. We first outline the general strategy and then consider the diagrams one by one. The results are tabulated in subsect. 6.8.

### *6.1 Preliminaries*

The diagrams contributing to $\Gamma^{(2,0,0)}(p,-p)^{ab}_{\mu\nu}$ at two-loop order are listed in fig. 2. In some cases the diagrams have been given a qualified label $i.j$, where $i$ denotes the type of the diagram and $j$ counts the diagrams of the given type. The total amplitude is then given by

$$\nu^{(2)}(p)\Big|_{\lambda_0=1} = \sum_{i=5}^{35} \nu_i(p), \tag{6.1}$$

where $\nu_i(p)$ is the sum of the contributions of the diagrams of type $i$.

In the continuum limit $p \to 0$ the amplitudes $\nu_i(p)$ may be expanded in a series of the form

$$\hat{p}^2\nu_i(p) = c_{0,i} + c_{1,i}\sum_{\mu=0}^{3}\frac{p_\mu^4}{p^2} + p^2\left\{c_{2,i}\left[\frac{\ln(p^2)}{(4\pi)^2}\right]^2 + c_{3,i}\frac{\ln(p^2)}{(4\pi)^2} + c_{4,i}\right\} + \mathrm{O}(p^4). \tag{6.2}$$

The coefficients $c_{n,i}$ have a simple dependence on $N$ given by

$$c_{n,i} = c_{n,i}^{(0)}/N^2 + c_{n,i}^{(1)} + c_{n,i}^{(2)}N^2. \tag{6.3}$$

The second term in the expansion (6.2) is of order $p^2$, but since it is not invariant under continuous Lorentz transformations, we expect that these terms cancel in the sum of all diagrams.

As already mentioned in sect. 4 the integrands of the diagrams are obtained by computer algebra. The general procedure to extract the coefficients $c_{n,i}$ from the corresponding Feynman integrals is as follows. We first expand the numerator of the integrand in powers of $p$. In most cases it is sufficient to work out the expansion up to second order, the remainder being negligible in the continuum limit. The terms are then classified according to their behaviour for small loop momenta. Terms proportional to sufficiently high powers of the loop momenta are easily dealt with. The expansion of the integral for $p \to 0$ may in these cases be obtained simply by expanding the integrand up to second order of $p$. One then ends up with a linear combination of integrals of the type

$$\int_{-\pi}^{\pi} \frac{\mathrm{d}^4 k}{(2\pi)^4} \frac{\mathrm{d}^4 q}{(2\pi)^4} \frac{P(k,q,r)}{(\hat{k}^2)^n (\hat{q}^2)^m (\hat{r}^2)^l}, \qquad r = -k - q, \tag{6.4}$$

where $P(k,q,r)$ is a polynomial in the sines and cosines of the momenta $k, q, r$ and $n, m, l$ are positive integers. Such integrals can be computed straightforwardly using the coordinate space methods described in ref.[3].

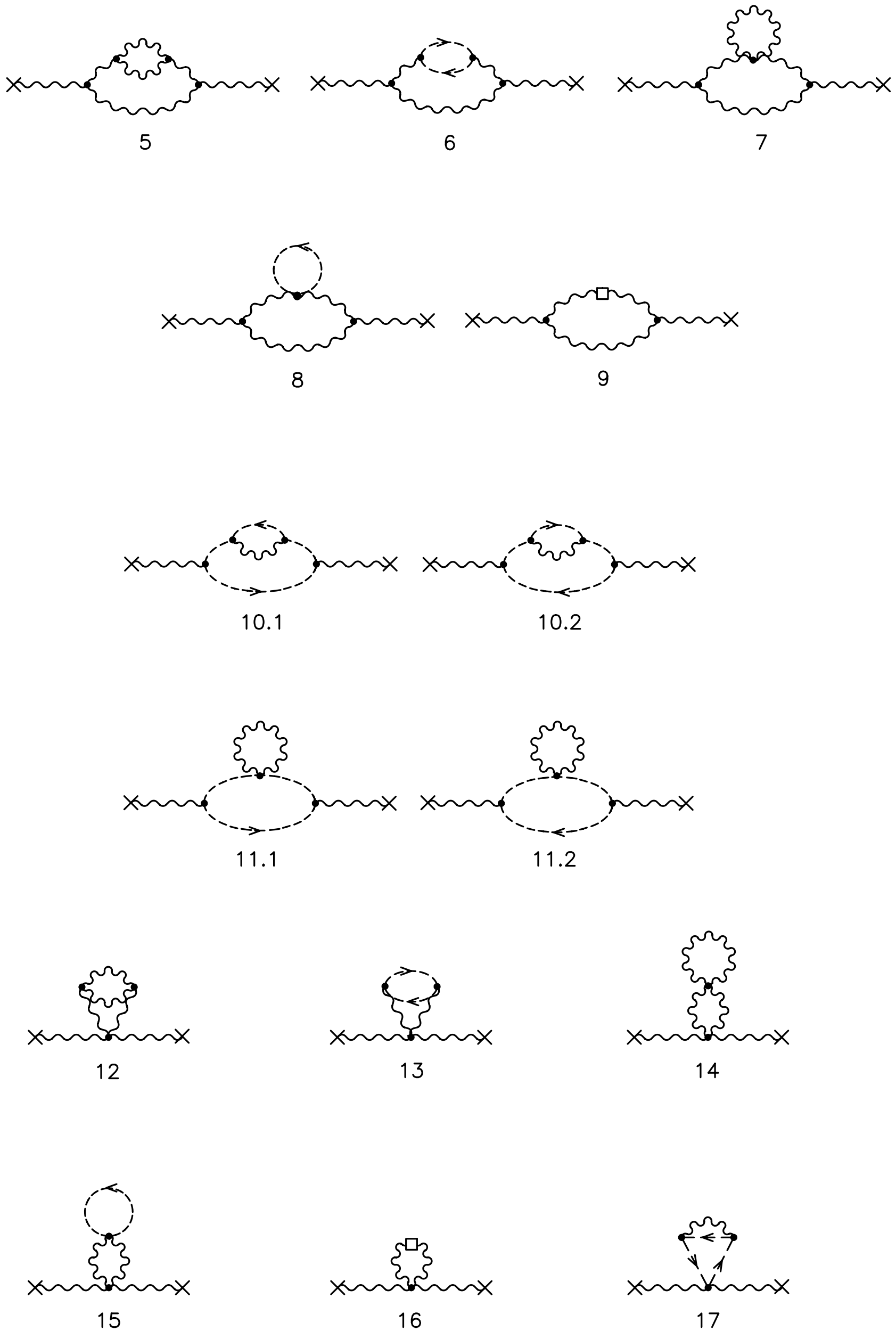

Fig. 2. Two-loop diagrams contributing to the background field 2-point function

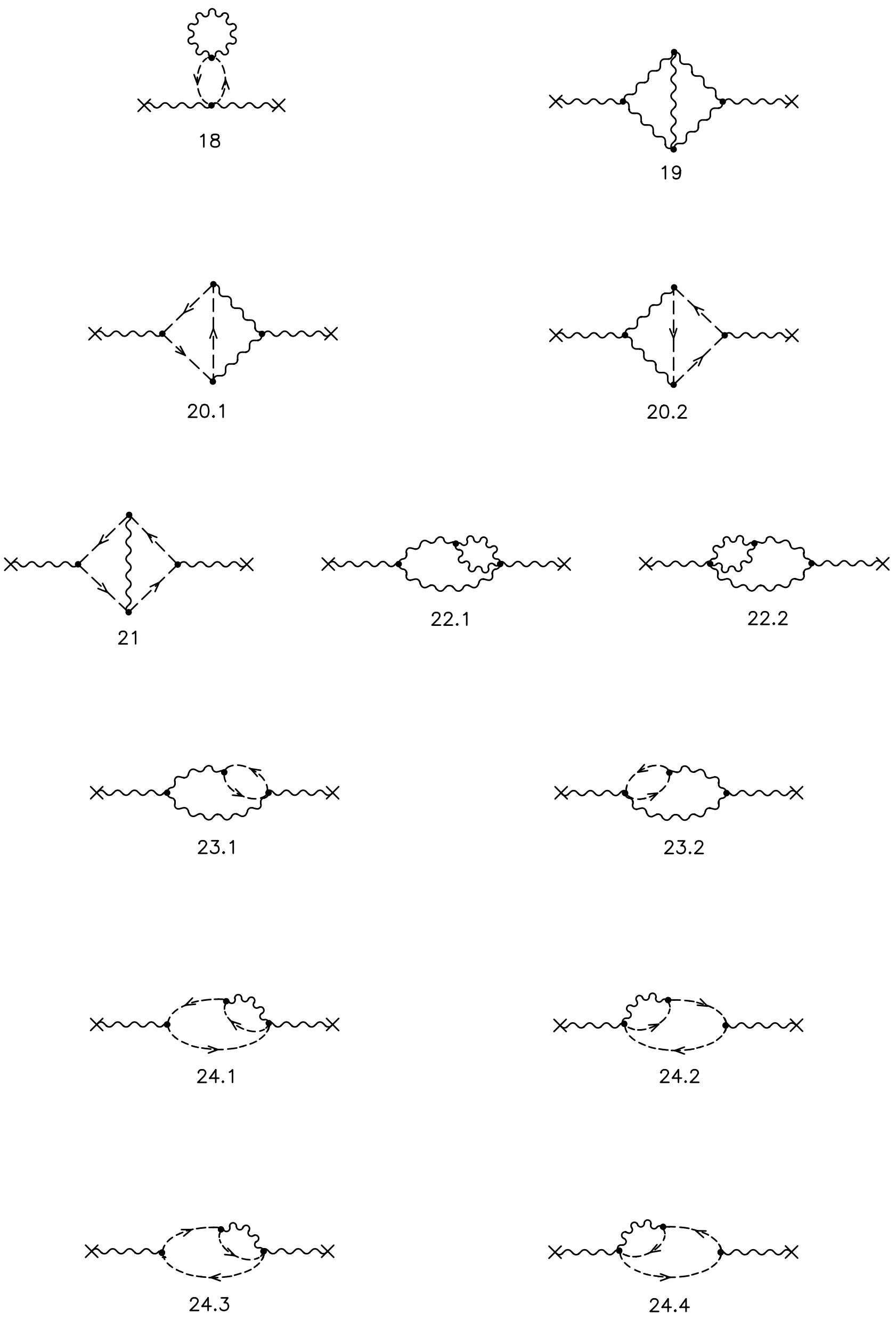

Fig. 2. (continued)

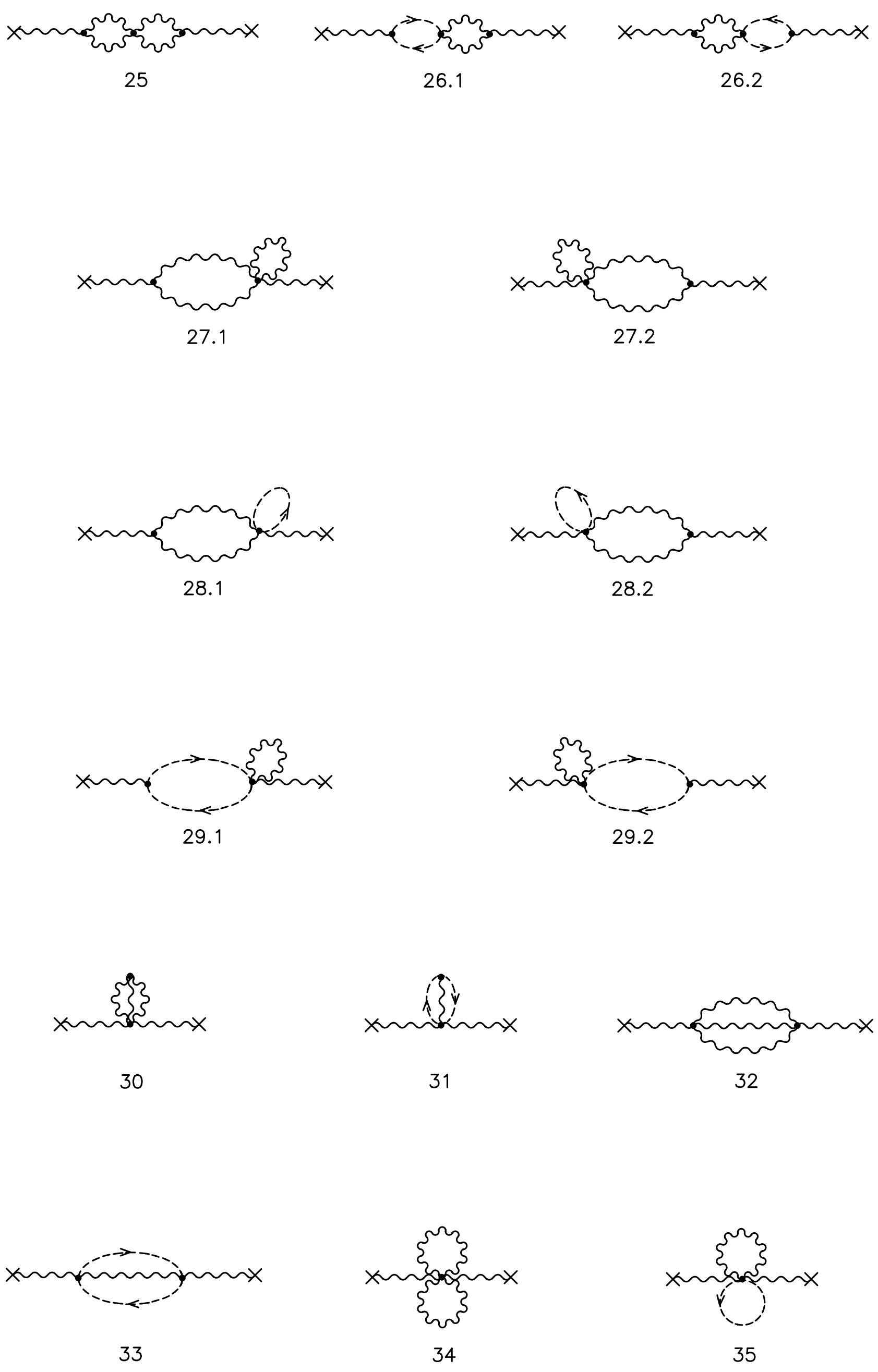

Fig. 2. (final page)

The terms left after this must be treated individually and are eventually reduced to the special one- and two-loop integrals listed in appendix B and C and integrals of the type (6.4). As in the case of the one-loop amplitudes considered in sect. 5, all algebraic manipulations of the integrands are programmable. The evaluation of the integrals of the type (6.4) (of which we had on the order of one hundred in our list) and those defined in appendix B and C is the only place where a numerical computation is required.

In the following subsections a few further indications of how to proceed in the various cases are given. It is not our aim to describe all manipulations in full detail. We just discuss some of the key steps that have been taken, hoping that this will be helpful to anyone interested to compute two-loop lattice Feynman integrals.

*6.2 Factorizable diagrams* ($i = 25 - 29, 34, 35$)

These diagrams can be written as sums of products of one-loop integrals which can be computed following the lines of sect. 5. In the continuum limit all these integrals can be expressed analytically in terms of the constants $P_1$ and $P_2$.

*6.3 Bigmac diagrams* ($i = 30 - 33$)

The integrals in this class are of the form

$$\hat{p}^2 \nu_i(p) = \int_{-\pi}^{\pi} \frac{\mathrm{d}^4 k}{(2\pi)^4} \frac{\mathrm{d}^4 q}{(2\pi)^4} \frac{F_i(p,k,q)}{\hat{k}^2 \hat{q}^2 \hat{r}^2}, \tag{6.5}$$

where $r = -k - q$ if $i = 30, 31$ and $r = -k - q - p$ otherwise. The numerator $F_i(p,k,q)$ is a polynomial in the sines and cosines of the momenta.

For $i = 30, 31$ the expansion of the integrals in the continuum limit $p \to 0$ is obtained simply by expanding the numerator of the integrand up to second order in $p$. As a consequence of the lattice symmetries, the resulting expression,

$$f(k,q) + \sum_{\mu=0}^{3} p_\mu f_\mu(k,q) + \sum_{\mu,\nu=0}^{3} p_\mu p_\nu f_{\mu\nu}(k,q), \tag{6.6}$$

may be replaced by

$$f(k,q) + \tfrac{1}{4}p^2 \sum_{\mu=0}^{3} f_{\mu\mu}(k,q). \tag{6.7}$$

One is then left with integrals of the type (6.4).

Before the integration is performed it is worthwhile to simplify the integrand as far as possible. This may be achieved by rewriting the numerator of the integrand as a polynomial in $\hat{k}_\mu^2$, $\hat{q}_\nu^2$, $\mathring{k}_\rho$ and $\mathring{q}_\sigma$. Using the trigonometric identities (4.14) and

$$\mathring{k}_\mu \mathring{q}_\mu = \tfrac{1}{2}(\hat{r}_\mu^2 - \hat{k}_\mu^2 - \hat{q}_\mu^2) + \tfrac{1}{4}\hat{k}_\mu^2 \hat{q}_\mu^2 \tag{6.8}$$

(with $r = -k - q$), all terms involving $\mathring{k}_\rho$ and $\mathring{q}_\sigma$ can be eliminated. The resulting expression is a sum of monomials in $\hat{k}_\mu^2$, $\hat{q}_\nu^2$ and $\hat{r}_\rho^2$. As it turns out there are just a few such terms left, and these are easily worked out using the coordinate space method.

In the case of the diagrams no. $i = 32, 33$, we expand $F_i(p,k,q)$ up to order $p^2$. Higher order terms are negligible since the scalar bigmac integral discussed in appendix C is absolutely convergent at $p = 0$. The numerator then is of the form (6.6).

The terms of order $p^0$ may be integrated as follows. We first split the integral in three pieces by substituting

$$f(k,q) = f(0,0) + [f(k,0) - f(0,0)] + [f(k,q) - f(k,0)]. \tag{6.9}$$

The first integral is proportional to the scalar bigmac integral (C.1). The second term is of order $k^2$ for small $k$. In this part of the integral we substitute $k \to k - p$ (so that $r \to -k - q$) and then obtain the desired expansion of the integral for $p \to 0$ by expanding the integrand to second order. The third part of the integral is treated in the same way, the substitution here being $q \to q - p$. In all cases one ends up with integrals of the type (6.4). As above the lattice symmetries and trigonometric identities should be used to simplify the integrands.

The integrals associated with the order $p^1$ and $p^2$ contributions to the numerator (6.6) are easy to work out. We simply insert the expansion of $1/\hat{r}^2$ to order $p^1$ and $p^0$, respectively. Again one is then left with integrals of the type (6.4).

*6.4 Tadpole diagrams* ($i = 12 - 18$)

These diagrams consist of a gluon or ghost loop with a self-energy insertion. The corresponding contributions $\nu_i(p)$ are formally given by

$$\hat{p}^2\nu_i(p) = \int_{-\pi}^{\pi} \frac{\mathrm{d}^4 k}{(2\pi)^4} (\hat{k}^2)^{-2} F_i(p,k), \qquad (6.10)$$

where $k$ is the momentum flowing through the loop. For $i = 16$ the function $F_i(p,k)$ is a polynomial in $\hat{p}_\mu^2$ and $\hat{k}_\nu^2$. In all other cases it is an integral over a second loop momentum $q$.

From the structure of the ghost-gluon vertices it follows that $F_{17}(p,k)$ and $F_{18}(p,k)$ are of order $k^2$ at $k = 0$. The integral (6.10) is thus well-defined in these cases. The same is not true for the other diagrams, where instead one may show that

$$F_i(p,k) = r_i F_{16}(p,k) + \mathrm{O}(k^2) \qquad (6.11)$$

for some constants $r_i$. Since the sum of all self-energy insertions is of order $k^2$ we have

$$\sum_{i=12}^{16} r_i = 0. \qquad (6.12)$$

The conclusion then is that while the integrals (6.10) do not separately exist for $i = 12 - 16$, their sum is perfectly well-defined. Taking eq.(6.12) into account, it is clear that the same result is obtained if we define $\nu_i(p)$ through the subtracted integral

$$\hat{p}^2\nu_i(p) = \int_{-\pi}^{\pi} \frac{\mathrm{d}^4 k}{(2\pi)^4} (\hat{k}^2)^{-2} \left[F_i(p,k) - r_i F_{16}(p,k)\right]. \qquad (6.13)$$

The expansion coefficients $c_{n,i}$ tabulated in subsect. 6.8 refer to this definition.

Since the denominators of the integrands do not depend on $p$, the expansion of the integrals in the continuum limit is obtained by expanding the numerators up to order $p^2$. The resulting integrals are of the type (6.4), or a subtracted variant of these integrals, such as

$$R_0 = \int_{-\pi}^{\pi} \frac{\mathrm{d}^4 k}{(2\pi)^4} (\hat{k}^2)^{-2} \int_{-\pi}^{\pi} \frac{\mathrm{d}^4 q}{(2\pi)^4} \sum_{\mu=0}^{3} \left\{ \frac{\hat{q}_\mu^2 \hat{r}_\mu^2}{\hat{q}^2 \hat{r}^2} - \frac{\hat{q}_\mu^4}{(\hat{q}^2)^2} \right\}. \qquad (6.14)$$

The coordinate space technique applies to integrals of this type as well.

*6.5 Eye diagrams* ($i = 22 - 24$)

The contributions from these diagrams are given by

$$\hat{p}^2 \nu_i(p) = \int_{-\pi}^{\pi} \frac{\mathrm{d}^4 k}{(2\pi)^4} \frac{\mathrm{d}^4 q}{(2\pi)^4} \frac{F_i(p,k,q)}{\hat{k}^2 \hat{s}^2 \hat{q}^2 \hat{r}^2}, \tag{6.15}$$

where $s = k + p$ and $r = -k - q$. The numerator $F_i(p,k,q)$ is a polynomial in the sines and cosines of the momenta. As usual we first expand $F_i(p,k,q)$ up to second order in $p$, rewrite the resulting expression as a polynomial in $p_\mu$, $\hat{k}_\nu^2$ etc., and eliminate products involving $\mathring{k}_\mu$ and $\mathring{q}_\mu$ as far as possible.

Using trigonometric identities and reexpanding in powers of $p$, the terms of lowest order in the momenta can be written as a linear combination of $p^2$, $\hat{k}^2$, $\hat{s}^2$, $\hat{q}^2$ and $\hat{r}^2$ plus terms of higher order. The integrals associated with the second order terms in the linear combination reduce to the scalar eye integral discussed in appendix C or to integrals already treated before.

The remaining terms are of fourth (or higher) order in the momenta. We first consider the terms of order $p^0$. They come in the form of a polynomial $f(k,q,r)$ in $\hat{k}_\mu^2$, $\hat{q}_\nu^2$ and $\hat{r}_\rho^2$. The polynomial is invariant under the lattice symmetries and is of order $k^2$ for small $k$. If we introduce the function

$$g(k) = \frac{1}{\hat{k}^2} \int_{-\pi}^{\pi} \frac{\mathrm{d}^4 q}{(2\pi)^4} \frac{f(k,q,r)}{\hat{q}^2 \hat{r}^2}, \tag{6.16}$$

the integral to be computed may be written as

$$I(p) = \int_{-\pi}^{\pi} \frac{\mathrm{d}^4 k}{(2\pi)^4} \frac{g(k)}{\hat{s}^2}. \tag{6.17}$$

$g(k)$ is smooth for all $k$ that do not vanish modulo $2\pi$. Moreover it follows from the properties of $f(k,q,r)$ that

$$g(k) \underset{k \to 0}{=} g(0) + \mathrm{O}(k^2). \tag{6.18}$$

It is now straightforward to show that

$$I(p) = (1 + \tfrac{1}{8}p^2) \int_{-\pi}^{\pi} \frac{\mathrm{d}^4 k}{(2\pi)^4} \frac{g(k)}{\hat{k}^2} - \tfrac{1}{4}p^2 \int_{-\pi}^{\pi} \frac{\mathrm{d}^4 k}{(2\pi)^4} \sum_{\mu=0}^{3} \frac{\hat{k}_\mu^4 g(k)}{(\hat{k}^2)^3}$$

$$- \frac{p^2}{32\pi^2} g(0) + \mathrm{O}(p^3), \tag{6.19}$$

where use has been made of the lattice symmetries to simplify the expressions on the right hand side. The constant $g(0)$ is a one-loop integral while the other terms appearing in eq.(6.19) are integrals of the type (6.4).

The terms of first and second order in $p$ are treated similarly. Some of these terms may not involve any factor of $\hat{k}_\mu^2$. In these cases the function $g(k)$ is defined with a subtraction at $k = 0$. One then ends up with subtracted integrals of the type (6.4), while the contribution from the subtraction term reduces to a product of one-loop integrals.

*6.6 Ring diagrams* ($i = 5 - 11$)

The integrals associated with these diagrams are of the form

$$\hat{p}^2 \nu_i(p) = \int_{-\pi}^{\pi} \frac{\mathrm{d}^4 k}{(2\pi)^4} \frac{1}{(\hat{k}^2)^2 \hat{s}^2} F_i(p,k), \qquad s = k + p. \tag{6.20}$$

For $i = 9$ the function $F_i(p,k)$ is a polynomial in the sines and cosines of the momenta, while in all other cases it is an integral over a second loop momentum $q$. A subtraction must be applied to some of the integrals (6.20) to make them converge at $k = 0$ (cf. subsect. 6.4). Explicitly, for $i = 5 - 9$, eq.(6.20) is replaced by

$$\hat{p}^2 \nu_i(p) = \int_{-\pi}^{\pi} \frac{\mathrm{d}^4 k}{(2\pi)^4} \frac{1}{(\hat{k}^2)^2 \hat{s}^2} [F_i(p,k) - r_i F_9(p,k)], \tag{6.21}$$

where the constants $r_i$ are determined through

$$F_i(p,k) = r_i F_9(p,k) + \mathrm{O}(k^2). \tag{6.22}$$

The sum of the subtracted integrals is the same as the sum of the unsubtracted integrals.

The integrals no. $i = 7 - 9, 11$ factorize into products of one-loop integrals and so are easy to compute. The other diagrams may be worked out essentially following the steps taken in the case of the eye diagrams (subsect. 6.5). Some of the more difficult integrands are those with numerators of order $p^2$ having the lowest possible degree in the loop momenta. These terms can be reduced to the integrals $E_{\mu\nu}(p)$ and $\mathcal{E}_{\mu\nu}(p)$ discussed in appendix C.

*6.7 Diamond diagrams* $(i = 19 - 21)$

The loop momenta in these diagrams may be assigned such that the corresponding Feynman integrals are given by

$$\hat{p}^2 \nu_i(p) = \int_{-\pi}^{\pi} \frac{\mathrm{d}^4 k}{(2\pi)^4} \frac{\mathrm{d}^4 q}{(2\pi)^4} \frac{F_i(p,k,q)}{\hat{k}_+^2 \hat{k}_-^2 \hat{r}^2 \hat{q}_+^2 \hat{q}_-^2}, \tag{6.23}$$

where the numerator $F_i(p,k,q)$ is a polynomial in the sines and cosines of the momenta and

$$k_\pm = k \pm \tfrac{1}{2}p, \qquad q_\pm = q \pm \tfrac{1}{2}p, \qquad r = -k - q. \tag{6.24}$$

Considering the symmetries of the denominator of the integrand, we may assume that $F_i(p,k,q)$ is invariant under the transformation $k,q \to -k,-q$.

In the continuum limit $p \to 0$, the scalar diamond integral is power-counting convergent and one may show that [18]

$$\int_{-\pi}^{\pi} \frac{\mathrm{d}^4 k}{(2\pi)^4} \frac{\mathrm{d}^4 q}{(2\pi)^4} \frac{1}{\hat{k}_+^2 \hat{k}_-^2 \hat{r}^2 \hat{q}_+^2 \hat{q}_-^2} = \frac{6}{(4\pi)^4} \zeta(3)/p^2 + \mathrm{O}(1).$$

A simple consequence of this observation is that the remainder in the expansion

$$F_i(p,k,q) = f(k,q) + \sum_{\mu,\nu=0}^{3} p_\mu p_\nu f_{\mu\nu}(k,q) + \sum_{\mu,\nu,\rho,\sigma=0}^{3} p_\mu p_\nu p_\rho p_\sigma f_{\mu\nu\rho\sigma}(k,q) + \mathrm{O}(p^6) \tag{6.25}$$

makes a negligible contribution to the integrals (6.23) and so may be dropped. Moreover it follows that the contribution of the fourth order term is given by

$$\frac{6}{(4\pi)^4} \zeta(3) \sum_{\mu,\nu,\rho,\sigma=0}^{3} p_\mu p_\nu p_\rho p_\sigma f_{\mu\nu\rho\sigma}(0,0)/p^2 + \mathrm{O}(p^4). \tag{6.26}$$

We are thus left with the first two terms in eq.(6.25).

The terms of order $p^2$ are first classified according to their behaviour for small $k,q$. Inspection shows that all terms are of second (or higher) order. Those which are of second degree may be reduced to integrals considered previously by applying simple trigonometric identities such as eq.(B.15) and

$$\hat{p}\mathring{k} = \tfrac{1}{2}(\hat{k}_+^2 - \hat{k}_-^2). \tag{6.27}$$

The terms of fourth and higher degree can be expanded straightforwardly and reduce to integrals of the type (6.4) or subtracted integrals of this type.

It turns out that the order $p^0$ terms may all be written in the form

$$f(k,q) = (\mathring{k}\mathring{q})g(k,q,r), \tag{6.28}$$

where $g(k,q,r)$ is a polynomial in $\hat{k}_\mu^2$, $\hat{q}_\nu^2$ and $\hat{r}_\rho^2$. The corresponding integral reduces to a product of one-loop integrals if $g(k,q,r)$ is proportional to $\hat{r}^2$. When $g(k,q,r)$ is a polynomial of lower degree 3 (or more) in both $\hat{k}$ and $\hat{q}$, the expansion of the integral is obtained by expanding the denominator of the integrand up to second order in $p$.

Up to perhaps a substitution $k,q \to q,k$ (which is a symmetry of the denominator of the integrand), the terms $g(k,q,r)$ left over then are

$$\hat{k}^2, \quad (\hat{k}^2)^2 \quad \text{and} \quad \sum_{\mu=0}^{3} \hat{k}_\mu^2 \hat{r}_\mu^2. \tag{6.29}$$

In the first two cases the integral may be reduced to integrals discussed before by inserting the identity (B.15). In the last case we expand the denominator after subtracting the integral over $k$ at $p = q = 0$ and end up with (subtracted) integrals of the type (6.4).

*6.8 Results*

We now list the coefficients $c_{n,i}$ appearing in the asymptotic expansion (6.2). The coefficients $c_{2,i}$ and $c_{3,i}$ of the logarithms can be given analytically in terms of the constants $P_1$ and $P_2$ introduced in appendix B (table 1). The coefficients $c_{1,i}$ vanish for all $i$ except for $i = 5, 7$, where we obtain

$$c_{1,5} = -c_{1,7} = \frac{N^2}{48\pi^2}\left(P_1 - \frac{1}{8}\right). \tag{6.30}$$

As anticipated in subsect. 6.1 the associated terms in the expansion (6.2) cancel after summing over all diagrams.

It remains to tabulate the coefficients $c_{0,i}$ and $c_{4,i}$. We begin by noting that the only non-zero contributions proportional to $1/N^2$ to these coefficients are

$$c_{4,14}^{(0)} = c_{4,32}^{(0)} = -2c_{4,34}^{(0)} = \tfrac{1}{64}. \tag{6.31}$$

Table 1. Coefficients $c_{2,i}$ and $c_{3,i}$

| $i$ | $c_{2,i}/N^2$ | $c_{3,i}/N^2$ |
|---|---|---|
| 5 | $\frac{15}{4}$ | $\frac{85}{36}P_1 - \frac{15}{2}P_2 - \frac{10}{9\pi^2} - \frac{5}{32}$ |
| 6 | $\frac{5}{12}$ | $\frac{5}{72}P_1 - \frac{5}{6}P_2 - \frac{25}{144\pi^2}$ |
| 7 | 0 | $-\frac{35}{12}P_1 - \frac{5}{32} + \frac{5}{8}N^{-2}$ |
| 10 | $\frac{1}{6}$ | $\frac{1}{24}P_1 - \frac{1}{3}P_2 - \frac{1}{12\pi^2}$ |
| 11 | 0 | $-\frac{1}{18}P_1$ |
| 19 | $\frac{27}{8}$ | $\frac{125}{144}P_1 - \frac{27}{4}P_2 - \frac{1213}{1152\pi^2}$ |
| 20 | $-\frac{1}{4}$ | $-\frac{23}{288}P_1 + \frac{1}{2}P_2 + \frac{43}{576\pi^2}$ |
| 21 | $\frac{1}{24}$ | $\frac{1}{96}P_1 - \frac{1}{12}P_2 - \frac{5}{128\pi^2}$ |
| 22 | $-\frac{9}{8}$ | $-\frac{119}{48}P_1 + \frac{9}{4}P_2 + \frac{9}{128\pi^2} + \frac{7}{32}$ |
| 23 | $-\frac{5}{24}$ | $\frac{5}{12}P_2 + \frac{7}{128\pi^2}$ |
| 24 | $-\frac{1}{6}$ | $-\frac{1}{24}P_1 + \frac{1}{3}P_2 + \frac{3}{32\pi^2}$ |
| 25 | $-6$ | $-\frac{13}{8}P_1 + 12P_2 + \frac{3}{2\pi^2} + \frac{3}{64}$ |
| 27 | 0 | $\frac{91}{24}P_1 + \frac{3}{64} - \frac{5}{8}N^{-2}$ |
| 29 | 0 | $\frac{1}{18}P_1$ |
| 32 | 0 | $\frac{3}{16\pi^2}$ |
| 33 | 0 | $-\frac{1}{48\pi^2}$ |

$c_{2,i} = c_{3,i} = 0$ for all $i$ not appearing in the first column

[cf. eq.(6.3)]. The contributions proportional to $N^0$ are listed in table 2. The constants $Q_1$ and $Q_2$ appearing there have been introduced in sect. 3 of ref.[3]. Their numerical values are

$$Q_1 = 0.0423063684(1), \tag{6.32}$$

$$Q_2 = 0.054623978180(1). \tag{6.33}$$

In table 3 we finally list the contributions proportional to $N^2$. Here only nu-

Table 2. Coefficients $c_{0,i}^{(1)}$ and $c_{4,i}^{(1)}$

| $i$ | $c_{0,i}^{(1)}$ | $c_{4,i}^{(1)}$ |
|---|---|---|
| 7 | $-P_1+\frac{1}{8}$ | $\frac{5}{96}P_1-\frac{5}{8}P_2-\frac{11}{192\pi^2}$ |
| 12 | 0 | $-\frac{1}{32}P_1^2-\frac{5}{192}P_1+\frac{5}{32}Q_1+\frac{1}{96}Q_2-\frac{1}{512}$ |
| 13 | 0 | $-\frac{5}{96}P_1^2+\frac{1}{96}P_1-\frac{1}{96}Q_1$ |
| 14 | $\frac{1}{2}P_1-\frac{1}{16}$ | $\frac{1}{24}P_1^2-\frac{1}{8}P_1-\frac{3}{256}$ |
| 27 | $P_1-\frac{1}{8}$ | $-\frac{13}{192}P_1+\frac{5}{8}P_2+\frac{7}{96\pi^2}$ |
| 30 | 0 | $\frac{3}{8}P_1^2-\frac{1}{48}P_1-\frac{5}{48}Q_1-\frac{1}{144}Q_2+\frac{1}{768}$ |
| 32 | 0 | $-\frac{1}{192}$ |
| 34 | $-\frac{1}{2}P_1+\frac{1}{16}$ | $-\frac{1}{4}P_1^2+\frac{13}{96}P_1+\frac{1}{512}$ |

$c_{0,i}^{(1)}=c_{4,i}^{(1)}=0$ for all $i$ not appearing in the first column

merical values are quoted, because the analytic expressions for the coefficients in terms of a basis of integrals of the type (6.4) would be too lengthy.

To guarantee the correctness of our results, all diagrams have been evaluated by both authors, using independent sets of programs. This includes the computation of the Feynman integrands, the expansion of the associated integrals for $p\to 0$ and, finally, the numerical calculation of the integrals of the type (6.4).

As a further check we note that the sum of the quadratically divergent terms vanishes,

$$\sum_{i=5}^{35} c_{0,i}=0, \tag{6.34}$$

as it should be. This can be verified from tables 2 and 3, but we have actually been able to establish the cancellation of these terms algebraically. The coefficients of the logarithms also add up to the expected values. Our result for the amplitude $\nu^{(2)}(p)$ then is

$$\nu^{(2)}(p)\Big|_{\lambda_0=1}=-\frac{N^2}{32\pi^4}\ln(p^2)+c_0/N^2+c_1+c_2N^2+\mathrm{O}(p^2), \tag{6.35}$$

Table 3. Coefficients $c_{0,i}^{(2)}$ and $c_{4,i}^{(2)}$

| $i$ | $c_{0,i}^{(2)} \times (4\pi)^2$ | $c_{4,i}^{(2)} \times (4\pi)^2$ |
|---|---|---|
| 5 | $-0.04037441(4)$ | $0.02348922(2)$ |
| 6 | $0.12388806(3)$ | $0.07287386(1)$ |
| 7 | $4.59937175(1)$ | $1.97571361(1)$ |
| 10 | $-0.09574219(1)$ | $0.01916905(1)$ |
| 11 | $-0.32549037(1)$ | $0.02049418(1)$ |
| 12 | $-0.09963952(2)$ | $-0.07789705(1)$ |
| 13 | $-0.07850005(1)$ | $0.03280987(1)$ |
| 14 | $-2.29968587(1)$ | $2.30107035(1)$ |
| 17 | $0.08108773(1)$ | $0$ |
| 18 | $0.16274518(1)$ | $0$ |
| 19 | $0.24305324(1)$ | $0.17668960(2)$ |
| 20 | $-0.00054087(1)$ | $-0.00823993(1)$ |
| 21 | $-0.02738405(1)$ | $0.00751124(1)$ |
| 22 | $-0.20607863(2)$ | $0.42631335(4)$ |
| 23 | $-0.09023515(1)$ | $-0.05439543(1)$ |
| 24 | $0.08461788(1)$ | $-0.02030710(1)$ |
| 25 | $0.07074594(1)$ | $-0.37533511(1)$ |
| 27 | $-4.74086363(1)$ | $-1.91757176(1)$ |
| 29 | $0.32549037(1)$ | $-0.02049418(1)$ |
| 30 | $-2.16944115(1)$ | $-0.09039338(1)$ |
| 31 | $0.23298215(1)$ | $0$ |
| 32 | $2.27248046(1)$ | $0.11846333(1)$ |
| 33 | $-0.23017351(1)$ | $0.00658707(1)$ |
| 34 | $2.37043182(1)$ | $-1.44106163(1)$ |
| 35 | $-0.16274518(1)$ | $0$ |

$c_{0,i}^{(2)} = c_{4,i}^{(2)} = 0$ for all $i$ not appearing in the first column

where the constants $c_j = \sum_{i=5}^{35} c_{4,i}^{(j)}$ are given by

$$c_0 = 3/128, \tag{6.36}$$

$$c_1 = -0.016544619540(4), \tag{6.37}$$

$$c_2 = 0.0074438722(2). \tag{6.38}$$

Note that to obtain the numerical values above, the algebraic results for the diagrams were summed before numerical evaluation.

## 7. Final results

Combining eqs.(2.5),(2.6) and (2.17)–(2.19) with our results for the background field 2-point function and the gluon propagator, we finally obtain

$$d_1(s) = -\frac{11N}{6\pi}\ln(s) - \frac{\pi}{2N} + k_1 N, \tag{7.1}$$

$$d_2(s) = d_1(s)^2 - \frac{17N^2}{12\pi^2}\ln(s) + \frac{3\pi^2}{8N^2} + k_2 + k_3 N^2, \tag{7.2}$$

where

$$k_1 = 2.135730074078457(2), \tag{7.3}$$

$$k_2 = -2.8626215972(6), \tag{7.4}$$

$$k_3 = 1.24911585(3). \tag{7.5}$$

The coefficients of $\ln(s)$ in eqs.(7.1) and (7.2) are of course directly related to the first two universal coefficients of the Callan-Symanzik $\beta$–function. The one-loop result was first obtained by A. and P. Hasenfratz [17]. It was subsequently reproduced using the background field technique in refs.[4–6].

Finally we note that the constant $k_2$ appearing above also enters the expansion

$$P = 1 - \frac{(N^2-1)}{2N}\pi\alpha_0 + (N^2-1)\left\{k_2 + \frac{5\pi^2}{24} + \frac{\pi^2}{8N^2}\right\}\alpha_0^2 + \dots \tag{7.6}$$

of the (normalized) plaquette expectation value $P$.

We are indebted to Anton van de Ven for performing various checks on our calculations, particularly of the results quoted in sect. 3 for the dimensionally regularized theory. We also thank Mrs. R. Heininger for producing the figures.

## Appendix A

We here list the non-vanishing 2- and 3-point vertices defined in subsect. 4.1. The function $P_{\mu\nu}(p)$ appearing in the expressions below is given by

$$P_{\mu\nu}(p) = \delta_{\mu\nu}\hat{p}^2 - \hat{p}_\mu\hat{p}_\nu, \tag{A.1}$$

where $\hat{p}_\mu = 2\sin(p_\mu/2)$.

We first write down all 2-point vertices. The inverse of the gluon and ghost field propagators are

$$V^{(0,2,0)}(p,-p)^{ab}_{\mu\nu} = \delta^{ab}\left\{P_{\mu\nu}(p) + \lambda_0\hat{p}_\mu\hat{p}_\nu\right\}, \tag{A.2}$$

$$V^{(0,0,1)}(p,-p)^{ab} = \delta^{ab}\hat{p}^2. \tag{A.3}$$

For the vertices with background field external legs one finds

$$V^{(2,0,0)}(p,-p)^{ab}_{\mu\nu} = g_0^{-2}\delta^{ab}P_{\mu\nu}(p), \tag{A.4}$$

$$V^{(1,1,0)}(p,-p)^{ab}_{\mu\nu} = g_0^{-1}\delta^{ab}P_{\mu\nu}(p), \tag{A.5}$$

and the 2-point "measure" vertex is given by

$$V^{(2)}_{\rm m}(p,-p)^{ab}_{\mu\nu} = \tfrac{1}{12}g_0^2N\delta^{ab}\delta_{\mu\nu}. \tag{A.6}$$

This completes the list of 2-point vertices.

We now tabulate the 3-point vertices with in-going momenta $p,q,r$ such that $p+q+r=0$. For the 3-point gluon vertex one obtains

$$V^{(0,3,0)}(p,q,r)^{abc}_{\mu\nu\rho} = ig_0f^{abc}\big\{\delta_{\mu\nu}(\widehat{p-q})_\rho\cos(r_\mu/2)$$
$$+\,\delta_{\nu\rho}(\widehat{q-r})_\mu\cos(p_\nu/2) + \delta_{\rho\mu}(\widehat{r-p})_\nu\cos(q_\rho/2)\big\}, \tag{A.7}$$

while the ghost-gluon vertex assumes the form

$$V^{(0,1,1)}(p,q,r)^{abc}_{\mu} = ig_0 f^{abc}\hat{q}_{\mu}\cos(r_{\mu}/2). \tag{A.8}$$

In these equations $f^{abc}$ denotes the SU($N$) structure constants as defined in appendix A of ref.[2]. The remaining 3-point vertices are

$$V^{(3,0,0)}(p,q,r)^{abc}_{\mu\nu\rho} = g_0^{-3}V^{(0,3,0)}(p,q,r)^{abc}_{\mu\nu\rho}, \tag{A.9}$$

$$\begin{aligned} V^{(2,1,0)}(p,q,r)^{abc}_{\mu\nu\rho} = {} & g_0^{-2}V^{(0,3,0)}(p,q,r)^{abc}_{\mu\nu\rho} \\ & + \tfrac{1}{2}g_0^{-1}f^{abc}\left\{\delta_{\mu\rho}P_{\mu\nu}(q) - \delta_{\nu\rho}P_{\mu\nu}(p)\right\}, \end{aligned} \tag{A.10}$$

$$\begin{aligned} V^{(1,2,0)}(p,q,r)^{abc}_{\mu\nu\rho} = {} & g_0^{-1}V^{(0,3,0)}(p,q,r)^{abc}_{\mu\nu\rho} \\ & + \tfrac{1}{2}f^{abc}\left\{\delta_{\mu\rho}P_{\nu\rho}(q) - \delta_{\mu\nu}P_{\nu\rho}(r)\right\} \\ & + i\lambda_0 f^{abc}\left\{\delta_{\mu\rho}\hat{q}_{\nu}\mathrm{e}^{-iq_{\mu}/2} - \delta_{\mu\nu}\hat{r}_{\rho}\mathrm{e}^{-ir_{\mu}/2}\right\}, \end{aligned} \tag{A.11}$$

$$V^{(1,0,1)}(p,q,r)^{abc}_{\mu} = if^{abc}(\widehat{q-r})_{\mu}. \tag{A.12}$$

## Appendix B

The integrals listed below can be expressed in terms of two constants,

$$P_1 = \int_{-\pi}^{\pi}\frac{\mathrm{d}^4k}{(2\pi)^4}\,\frac{1}{\hat{k}^2}, \tag{B.1}$$

$$P_2 = \lim_{m\to 0}\left\{\frac{1}{(4\pi)^2}\ln(m^2) + \int_{-\pi}^{\pi}\frac{\mathrm{d}^4k}{(2\pi)^4}\,\frac{1}{(\hat{k}^2+m^2)^2}\right\}. \tag{B.2}$$

Their numerical values

$$P_1 = 0.15493339023106021(1), \tag{B.3}$$

$$P_2 = 0.02401318111946489(1), \tag{B.4}$$

have been determined in ref.[3].

In the course of our computations a small number of one-loop integrals with zero external momentum occurred. Besides the basic integrals (B.1) and (B.2) the list of these integrals includes

$$\int_{-\pi}^{\pi} \frac{\mathrm{d}^4 k}{(2\pi)^4} \sum_{\mu=0}^{3} \frac{\hat{k}_\mu^4}{(\hat{k}^2)^2} = 1 - 4P_1, \tag{B.5}$$

$$\int_{-\pi}^{\pi} \frac{\mathrm{d}^4 k}{(2\pi)^4} \sum_{\mu=0}^{3} \frac{\hat{k}_\mu^4}{(\hat{k}^2)^3} = \frac{1}{2}P_1 - \frac{1}{8\pi^2}, \tag{B.6}$$

$$\int_{-\pi}^{\pi} \frac{\mathrm{d}^4 k}{(2\pi)^4} \sum_{\mu=0}^{3} \frac{\hat{k}_\mu^6}{(\hat{k}^2)^3} = 1 - 5P_1 - \frac{1}{2\pi^2}. \tag{B.7}$$

To establish eqs.(B.5)–(B.7) one makes use of the identity (4.14) and partial integration.

In the case of the integrals depending on the external momentum $p$ we are only interested in their asymptotic form in the continuum limit $p \rightarrow 0$. The scalar self-energy diagram

$$A(p) = \int_{-\pi}^{\pi} \frac{\mathrm{d}^4 k}{(2\pi)^4} \frac{1}{\hat{k}_+^2 \hat{k}_-^2}, \qquad k_\pm = k \pm \tfrac{1}{2}p, \tag{B.8}$$

has been worked out in ref.[3] using a momentum space subtraction method, the result being

$$A(p) = \frac{1}{(4\pi)^2}\left[-\ln(p^2) + 2\right] + P_2 + \mathrm{O}(p^2). \tag{B.9}$$

Here and below the error term $\mathrm{O}(p^n)$ stands for a remainder $R(p)$ such that $\lim_{p\rightarrow 0} R(p)/|p|^{n-\epsilon} = 0$ for all $\epsilon > 0$.

The momentum space subtraction technique may also be applied to the more complicated integral

$$C_{\mu\nu}(p) = \int_{-\pi}^{\pi} \frac{\mathrm{d}^4 k}{(2\pi)^4} \mathring{k}_\mu \mathring{k}_\nu \left\{ \frac{1}{\hat{k}_+^2 \hat{k}_-^2} - \frac{1}{(\hat{k}^2)^2} \right\} \tag{B.10}$$

[cf. eq.(4.13)]. The calculations can be simplified by noting that

$$\sum_{\mu=0}^{3} \hat{p}_\mu C_{\mu\nu}(p) = 0, \tag{B.11}$$

and the result then is

$$C_{\mu\nu}(p) = \left(\delta_{\mu\nu} p^2 - p_\mu p_\nu\right) \left\{ -\frac{1}{12} A(p) - \frac{1}{18(4\pi)^2} + \frac{1}{72} P_1 \right\} + \mathrm{O}(p^4). \tag{B.12}$$

To deduce the expansions

$$\int_{-\pi}^{\pi} \frac{\mathrm{d}^4 k}{(2\pi)^4} \frac{\hat{k}^2}{\hat{k}_+^2 \hat{k}_-^2} = P_1 + p^2 \left\{ -\frac{1}{4} A(p) + \frac{1}{32} P_1 \right\} + \mathrm{O}(p^4), \tag{B.13}$$

$$\int_{-\pi}^{\pi} \frac{\mathrm{d}^4 k}{(2\pi)^4} \frac{(\hat{k}^2)^2}{\hat{k}_+^2 \hat{k}_-^2} = 1 + \mathrm{O}(p^4), \tag{B.14}$$

we insert the identity

$$\hat{k}^2 = \tfrac{1}{2}(\hat{k}_+^2 + \hat{k}_-^2) - \tfrac{1}{4} p^2 + \tfrac{1}{8} \sum_{\mu=0}^{3} p_\mu^2 \hat{k}_\mu^2 + \mathrm{O}(p^4) \tag{B.15}$$

in the numerator of the integrands and expand in powers of $p$ after cancelling factors of $\hat{k}_\pm^2$.

We finally quote our result for three further one-loop integrals which were needed to evaluate some two-loop diagrams. They are given by

$$\int_{-\pi}^{\pi} \frac{\mathrm{d}^4 k}{(2\pi)^4} \frac{\mathring{k}_\mu}{\hat{k}^2 \hat{s}^2} = p_\mu \left\{ -\frac{1}{2} A(p) + \frac{1}{16} P_1 \right\} + \mathrm{O}(p^3), \tag{B.16}$$

$$\int_{-\pi}^{\pi} \frac{\mathrm{d}^4 k}{(2\pi)^4} \frac{\hat{k}_\mu^2}{(\hat{k}^2)^2 \hat{s}^2} = \frac{1}{2(4\pi)^2} \frac{p_\mu^2}{p^2} + \frac{1}{4} A(p) - \frac{1}{8(4\pi)^2} + \mathrm{O}(p^2), \tag{B.17}$$

$$\begin{aligned} \int_{-\pi}^{\pi} \frac{\mathrm{d}^4 k}{(2\pi)^4} \frac{\mathring{k}_\mu \mathring{k}_\nu}{(\hat{k}^2)^2 \hat{s}^2} = {} & \frac{1}{2(4\pi)^2} \frac{p_\mu p_\nu}{p^2} \\ & + \delta_{\mu\nu} \left\{ \frac{1}{4} A(p) - \frac{1}{32} P_1 \right\} + \mathrm{O}(p^2), \end{aligned} \tag{B.18}$$

where $s = k + p$.

## Appendix C

The two-loop integrals listed here have been worked out using a mixture of momentum and coordinate space techniques of the sort discussed in sect. 4 of ref.[3]. There are four basic integrals that must be evaluated in this way and thus require special attention. The simplest of them, already considered in ref.[3], is the scalar "bigmac" integral

$$D(p) = \int_{-\pi}^{\pi} \frac{\mathrm{d}^4 k}{(2\pi)^4} \frac{\mathrm{d}^4 q}{(2\pi)^4} \frac{1}{\hat{k}^2 \hat{q}^2 \hat{r}^2}, \qquad r = -k - q - p. \tag{C.1}$$

In the continuum limit $p \to 0$ the asymptotic expansion

$$D(p) = A_0 + p^2 \left[ \frac{1}{2(4\pi)^4} \ln(p^2) + A_1 \right] + \mathrm{O}(p^4) \tag{C.2}$$

holds, where

$$A_0 = 0.0040430548122(3), \tag{C.3}$$

$$A_1 = -0.00007447695(1). \tag{C.4}$$

The integral corresponding to the scalar "eye" diagram,

$$E(p) = \int_{-\pi}^{\pi} \frac{\mathrm{d}^4 k}{(2\pi)^4} \frac{1}{\hat{k}^2 \hat{s}^2} A(k), \qquad s = k + p, \tag{C.5}$$

has also been worked out in ref.[3] (the function $A(k)$ is discussed in appendix B). In the continuum limit it is given by

$$E(p) = \frac{1}{2(4\pi)^4} \left[\ln(p^2)\right]^2 - \frac{1}{(4\pi)^2} \left[ P_2 + \frac{3}{(4\pi)^2} \right] \ln(p^2) + X_1 + \mathrm{O}(p^2), \tag{C.6}$$

where, for the constant $X_1$, the value

$$X_1 = 0.001123187953(1) \tag{C.7}$$

has been obtained.

Two further integrals, related to the "ring" diagrams, are defined through

$$E_{\mu\nu}(p) = \int_{-\pi}^{\pi} \frac{\mathrm{d}^4 k}{(2\pi)^4} \frac{\mathring{k}_\mu \mathring{k}_\nu}{(\hat{k}^2)^2 \hat{s}^2} A(k), \qquad s = k + p, \tag{C.8}$$

$$\mathcal{E}_{\mu\nu}(p) = \int_{-\pi}^{\pi} \frac{\mathrm{d}^4 k}{(2\pi)^4} \frac{1}{(\hat{k}^2)^2 \hat{s}^2} \cos(\tfrac{1}{2}k_\mu)\cos(\tfrac{1}{2}k_\nu) C_{\mu\nu}(k) \tag{C.9}$$

[cf. eq.(4.13)]. Note that the function $C_{\mu\nu}(k)$ (which was introduced in appendix B) is not invariant under shifts of the components of $k$ by integer multiples of $2\pi$, but the integrand in eq.(C.9) is. The asymptotic expressions for these integrals have the tensor structure

$$E_{\mu\nu}(p) = \delta_{\mu\nu}\, E_1(p) + \frac{p_\mu p_\nu}{p^2}\, E_2(p) + \mathrm{O}(p^2), \tag{C.10}$$

$$\mathcal{E}_{\mu\nu}(p) = \delta_{\mu\nu}\, \mathcal{E}_1(p) + \frac{p_\mu p_\nu}{p^2}\, \mathcal{E}_2(p) + \mathrm{O}(p^2). \tag{C.11}$$

For the scalar amplitudes $E_1(p)$ and $\mathcal{E}_1(p)$ one finds

$$E_1(p) = \frac{1}{4}\Bigg\{ \frac{1}{2(4\pi)^4}\left[\ln(p^2)\right]^2 - \frac{1}{(4\pi)^2}\left[P_2 + \frac{5}{2(4\pi)^2}\right]\ln(p^2) + X_2 \Bigg\}, \tag{C.12}$$

$$\mathcal{E}_1(p) = -\frac{1}{16}\Bigg\{ \frac{1}{2(4\pi)^4}\left[\ln(p^2)\right]^2 - \frac{1}{(4\pi)^2}\left[P_2 - \frac{1}{6}P_1 + \frac{23}{6(4\pi)^2}\right]\ln(p^2) + X_3 \Bigg\}, \tag{C.13}$$

with the constants $X_2$ and $X_3$ being given by

$$X_2 = 0.000583430648(4), \tag{C.14}$$

$$X_3 = 0.000403986926(4). \tag{C.15}$$

For the other amplitudes, $E_2(p)$ and $\mathcal{E}_2(p)$, one obtains

$$E_2(p) = \frac{1}{2(4\pi)^4}\left[-\ln(p^2) + \frac{3}{2}\right] + \frac{1}{2(4\pi)^2}P_2, \tag{C.16}$$

$$\mathcal{E}_2(p) = \frac{1}{24(4\pi)^4}\left[-\ln(p^2) + \frac{13}{6}\right] + \frac{1}{24(4\pi)^2}\left[P_2 - \frac{1}{6}P_1\right]. \tag{C.17}$$

## References

[1] M. Lüscher and P. Weisz, Two-loop relation between the bare lattice coupling and the $\overline{\mathrm{MS}}$ coupling in pure SU($N$) gauge theories, preprint MPI-PhT/95-5, hep-lat/9502001, to appear in Phys. Lett. B

[2] M. Lüscher and P. Weisz, Background field technique and renormalization in lattice gauge theory, preprint MPI-PhT/95-27, hep-lat/9504006

[3] M. Lüscher and P. Weisz, Coordinate space methods for the evaluation of Feynman diagrams in lattice field theories, preprint MPI-PhT/95-13, hep-lat/9502017, to appear in Nucl. Phys. B

[4] R. Dashen and D. Gross, Phys. Rev. D23 (1981) 2340

[5] A. Hasenfratz and P. Hasenfratz, Nucl. Phys. B193 (1981) 210

[6] A. Gonzalez-Arroyo and C. P. Korthals-Altes, Nucl. Phys. B205 (1982) 46

[7] L. F. Abbott, Nucl. Phys. B185 (1981) 189

[8] D. M. Capper and A. Maclean, Nucl. Phys. B203 (1982) 413

[9] W. A. Bardeen, A. Buras, D. W. Duke and T. Muta, Phys. Rev. D18 (1978) 3998

[10] E. Braaten and J. P. Leveille, Phys. Rev. D24 (1981) 1369

[11] R. K. Ellis, Perturbative corrections to universality and renormalization group behaviour, *in*: Gauge Theory on a Lattice: 1984, Proceedings of the Argonne National Laboratory Workshop, Eds. C. Zachos et al. (Argonne 1984).

[12] A. van de Ven, unpublished

[13] I. S. Gradshteyn and I. M. Ryzhik, Table of integrals, series and products, 4th edition (Academic Press, New York, 1965)

[14] M. Lüscher and P. Weisz, Nucl. Phys. B266 (1986) 309

[15] B. Allés, M. Campostrini, A. Feo and H. Panagopoulos, Nucl. Phys. B413 (1994) 553
[16] S. Capitani and G. Rossi, The use of SCHOONSCHIP and FORM in perturbative lattice calculations, preprint ROM2F/95/06, hep-lat/9504014
[17] A. Hasenfratz and P. Hasenfratz, Phys. Lett. 93B (1980) 165
[18] K. G. Chetyrkin and F. V. Tkachov, Nucl. Phys. B192 (1981) 159